\documentclass[twocolumn,prl,showpacs,amsmath,amssymb]{revtex4}

\usepackage{graphicx}
\usepackage{dcolumn}
\usepackage{bm}
\usepackage[all]{xy}

 \newcommand{\Dscr}{\mathcal{D}}

\newcommand{\Lscr}{\mathcal{L}}

\newcommand{\rmd}{{\mathrm{d}}}
\newcommand{\rmi}{{\mathrm{i}}}
\newcommand{\rme}{\mathrm{e}}

\newcommand{\Real}{\mathbb{R}}
\newcommand{\Complex}{\mathbb{C}}

\newcommand{\abs}[1]{\left\vert#1\right\vert}

\newcommand{\Tr}{\operatorname{Tr}}
\newcommand{\E}{\operatorname{\mathbb{E}}}

\begin{document}

\title{Feedback control of the fluorescence light squeezing}

\author{A.~Barchielli}
\altaffiliation[also: ]{INFN, Sezione di Milano}
\author{M.~Gregoratti}
\altaffiliation[also: ]{INFN, Sezione di Milano}
\author{M.~Licciardo}
\affiliation{Mathematics Department, Politecnico di Milano,
piazza Leonardo da Vinci 32, I-20133, Milano, Italy}

\date{\today}

\begin{abstract}
We consider a two-level atom stimulated by a coherent monochromatic laser and we study how to
enhance the squeezing of the fluorescence light and of the atom itself in the presence of a
Wiseman-Milburn feedback mechanism, based on the homodyne detection of a fraction of the
emitted light. Besides analyzing the effect of the control parameters on the squeezing
properties of the light and of the atom, we also discuss the relations among these. The
problem is tackled inside the framework of quantum trajectory theory.
\end{abstract}


\maketitle

Photo-detection theory in continuous time has been widely developed
\cite{Dav76,Bar90QO,BarPag96,Bar05} and applied, in particular, to the fluorescence light
emitted by a two-level atom stimulated by a coherent monochromatic laser
\cite{WisMil93,Bar05}. As well as various feedback schemes on the atom evolution, based on the
outcoming photocurrent, have been proposed \cite{WisM93,WW00}. However the introduction and
the analysis of feedback have been mainly focused on the control of the atom \cite{WW00}.

Here, we are interested not only in the atom, but also, and mainly, in the emitted light and
in employing control and feedback processes to enhance the squeezing properties of both
systems. The squeezing of the fluorescence light can be checked by homodyne detection and
spectral analysis of the output current \cite{WalM94,BGL08}. For these reasons we consider the
mathematical description of photo-detection based on quantum trajectories, as it is suitable
both to consistently compute the homodyne spectrum of fluorescence light, and to introduce
feedback and control in the mathematical formulation. We study how the squeezing depends on
the various control parameters, how feedback mechanisms can be successfully introduced and
which is the relationship among the squeezing properties of the quantum systems involved. We
consider only Markovian feedback schemes \`a la Wiseman-Milburn \cite{WisM93}, as they leave
the homodyne spectrum explicitly computable.

\paragraph{Two-level atom and atomic squeezing.}
Consider a two-level atom with Hilbert space $\Complex^2$ and lowering and rising
operators $\sigma_-$ and $\sigma_+$. Let us denote by $\vec\sigma$ the vector
$(\sigma_x,\sigma_y,\sigma_z)$ of the Pauli matrices. Let the
eigenprojectors of $\sigma_z$ be denoted by $P_+$ and $P_-$
and, for every angle $\phi$, let us introduce the unitary selfadjoint operator
\begin{equation*}
\sigma_\phi = \rme^{\rmi\phi}\,\sigma_- + \rme^{-\rmi\phi}\,\sigma_+
= \cos\phi\,\sigma_x+\sin\phi\,\sigma_y.
\end{equation*}
A state $\rho$ of the atom is represented by a point $\vec{x}$ in the Bloch sphere,
$\rho=\left(1+\vec{x}\cdot\vec\sigma\right)/2$, with $\vec{x}\in\Real^3$, $\abs{\vec{x}}\leq1$.

Walls and Zoller \cite{WallsZ81} suggested to define the squeezing of a two-level atom from
the Heisenberg-Robertson uncertainty relations for $\sigma_x$ and $\sigma_y$. By using the
equatorial component $\sigma_\bot$ of $\vec\sigma$ with minimum variance $\Delta
\sigma_\bot^2$, we say that the atomic state $\rho$ is squeezed if
\begin{equation*}
\Delta \sigma_\bot^2 = 1-(x^2+y^2)< \abs{\langle \sigma_z\rangle }=\abs{z}.
\end{equation*}
We call \emph{atomic squeezing parameter} of $\rho$ the quantity
\[
\mathrm{AS}_\rho= 1 - (x^2+y^2)-\abs{z},
\]
so that $\rho$ is squeezed if $\mathrm{AS}_\rho<0$. Thus
$\min_\rho \mathrm{AS}_\rho=-1/4$, attained by pure states with $x^2+y^2=3/4$
and $z^2=1/4$.

\paragraph{Detection and feedback scheme.}
We admit an open Markovian evolution for the atom, subjected to `dephasing' effects and to
interactions both with a thermal bath and with the electromagnetic field, via absorption and emission of photons. The atom is stimulated by a coherent
monochromatic laser and the emitted light is partially lost in the
\textit{forward channel} and partially gathered in two
\textit{side channels} for homodyne detection.

Let the free Hamiltonian of the atom be $\omega_0\sigma_z/2$, with $\omega_0>0$. Let the natural
line-width of the atom be $\gamma>0$, let the intensities of the dephasing and thermal effects
be $k_\rmd\geq 0$ and $\overline n\geq 0$, let the
stimulating laser have frequency $\omega>0$. Finally, let the Rabi frequency be $\Omega\geq0$
and let $\Delta\omega =\omega_0-\omega$ denote the detuning.

Let the fractions of light emitted in the forward and in the two side channels be
$|\alpha_0|^2$, $|\alpha_1|^2$, $|\alpha_2|^2$, respectively
($|\alpha_0|^2+|\alpha_1|^2+|\alpha_2|^2=1$); for $k=1,2$, we can say that $|\alpha_k|^2$ is
the efficiency of the detector $k$. Let the initial phase of the local oscillator in each
detector be $\vartheta_k$, included in the parameter $\alpha_k\in\mathbb{C}$ by setting
$\vartheta_k=\arg\alpha_k$. To change $\vartheta_k$ means to change the measuring apparatus.
Let the two homodyne photocurrents be $I_1$ and $I_2$.

We introduce a feedback scheme \`a la Wiseman-Milburn based on $I_1$. Assuming instantaneous
feedback, we modify the amplitude of the stimulating laser by adding a term
$g\,\rme^{-\rmi\omega t}\,I_1(t)/\sqrt\gamma$ proportional to $I_1$, with the same frequency
$\omega$ and with initial phase possibly different from that of the original laser. Let this
phase difference be $\varphi$.
\[ \xymatrix{
& & & &
\\
*+[F]{\txt{\begin{scriptsize}h. det.\end{scriptsize}}} \ar[u]^>>>{I_2(t)}& &
*+[F-:<3pt>]{\txt{\begin{scriptsize}atom\end{scriptsize}}}
\ar@{~>}[u]^>>>{\text{forward}}_>>>{\text{channel}}\ar@{~>}[rr]^{\text{side}}_{\text{channel
1}} \ar@{~>}[ll]_{\text{side}}^{\text{channel 2}} & & *+[F]{\txt{\begin{scriptsize}h.
det.\end{scriptsize}}} \ar@/^1pc/[dll]^<<<<<<<{I_1(t)}
\\
&& *+[F]{\txt{\begin{scriptsize}electromodulator\end{scriptsize}}} \ar@{~>}[u]& &
\\
&& \ar@{~} [u]_<<<{\text{laser}} & &}
\]

Then the atom has a Markovian evolution, whether we condition its state on continuous
monitoring of the photocurrents, or we do not. Let us call \textit{a priori} state $\eta_t$
the unconditioned one and let us call \textit{a posteriori} state $\rho_t$ the conditioned
one. Of course $\eta_t$ is the mean of $\rho_t$. Let us write the evolution equations in the
rotating frame, where they result to be time-homogeneous. Let us introduce first the
parameters $c=|g|\,|\alpha_0|/\sqrt\gamma\geq0$, and
$\Delta\omega_c=\Delta\omega+c\,\gamma\,|\alpha_1|\cos(\vartheta_1-\varphi)\in\Real$. The a
priori state $\eta_t$ is governed by the master equation $ \rmd\eta_t= \Lscr\eta_t\,\rmd t$,
where
\begin{eqnarray*}
\Lscr\rho &=& -\rmi\left[\frac{\Delta\omega_c}{2}\,\sigma_z +
\frac{\Omega}{2}\,\sigma_x\;,\;\rho\right] + \gamma k_\rmd
\left(\sigma_z\,\rho\,\sigma_z-\rho\right)
\\ &+&
\gamma\overline{n}\left(\sigma_+\,\rho\,\sigma_--\frac{1}{2}\left\{P_-\;,\;\rho\right\}\right)
\\
&+&\gamma(\overline{n}+1-|\alpha_1|^2) \left(\sigma_-\,\rho\,\sigma_+-\frac{1}{2}
\left\{P_+\;,\;\rho\right\}\right)
\\ &+&\gamma(\alpha_1\,\sigma_--\rmi
c\,\sigma_\varphi)\,\rho\,(\overline{\alpha}_1\,\sigma_++\rmi c\,\sigma_\varphi)
\\ &-&
\frac{\gamma}{2}\left\{\Big(|\alpha_1|^2-2c|\alpha_1|\sin(\vartheta_1-\varphi)\Big)P_+
+c^2\;,\;\rho\right\}.
\end{eqnarray*}
The a posteriori state $\rho_t$ is governed by the non-linear stochastic master
equation
\begin{eqnarray}\nonumber
\rmd\rho_t&=& \Lscr\rho_t\,\rmd t + \sqrt\gamma \Dscr[\alpha_1\,\sigma_--\rmi
c\,\sigma_\varphi]\rho_t\,\rmd
W_1(t)
\\ \label{posterior}
&&{}+ \sqrt\gamma \Dscr[\alpha_2\,\sigma_-]\rho_t\,\rmd W_2(t),
\end{eqnarray}
where $\Dscr[a]\rho=a\,\rho + \rho\,a^* - \rho \Tr\left[(a+a^*)\rho\right]$ for every matrix $a$,
and where $W_1$ and $W_2$ are two independent standard Wiener processes. The two homodyne
photocurrents are given by the generalized stochastic processes
\begin{equation}\label{currents}
I_k(t)=\sqrt\gamma|\alpha_k| \Tr\left[\sigma_{\vartheta_k}\,\rho_t\right] + \dot{W}_k(t).
\end{equation}

We suppose that $|\alpha_0|$ is assigned by experimental constraints and that the control
parameters are $\Omega$, $\Delta\omega$, $\vartheta_1$, $\vartheta_2$, $c$, $\varphi$ and,
eventually, $|\alpha_1|$ and $|\alpha_2|$. Of course, if $c=0$ there is no feedback action on
the atom, so that its a priori dynamics is independent of the measurement process, that is of
$\alpha_1$, $\alpha_2$ and $\varphi$. On the contrary, if $c>0$, then the a priori dynamics is
modified by the feedback loop and it depends also on $\alpha_1$, $\varphi$ and $c$.

In the Bloch sphere language $\Lscr$ is an affine map. Let its linear part be given by the
matrix  $-A$, where
\begin{eqnarray*}
A&=&\begin{pmatrix}a_{11}&a_{12}&0\\
a_{21}&a_{22}&\Omega\\
0&-\Omega&a_{33}\end{pmatrix},
\\
a_{11}&=&\gamma\Big(\frac{1}{2}+\overline{n}+2k_\rmd + 2c|\alpha_1|\cos\vartheta_1
\sin\varphi+2c^2\sin^2\varphi\Big),
\\
a_{12}&=&\Delta\omega_c -
\gamma\Big(c|\alpha_1|\cos(\vartheta_1+\varphi)+c^2\sin2\varphi\Big),
\\
a_{21}&=&-\Delta\omega_c -
\gamma\Big(c|\alpha_1|\cos(\vartheta_1+\varphi)+c^2\sin2\varphi\Big),
\\
a_{22}&=&\gamma\Big(\frac{1}{2}+\overline{n}+2k_\rmd - 2c|\alpha_1|\sin\vartheta_1
\cos\varphi+2c^2\cos^2\varphi\Big),
\\
a_{33}&=&\gamma\Big(1+2\overline{n} - 2c|\alpha_1|\sin(\vartheta_1-\varphi)+2c^2\Big).
\end{eqnarray*}
Apart from the exceptional case $\det A=0$, which occurs if and only if $k_\rmd=\overline
n=0$, $|\alpha_1|=1$, $2c\sin(\vartheta_1-\varphi)=1$, $\Omega\sin\vartheta_1=0$,
$\Delta\omega=-\gamma c\cos(\vartheta_1-\varphi)$, the a priori dynamics has a unique stable
stationary state $\eta_\mathrm{eq}=\left(1+\vec{x}_\mathrm{eq}\cdot\vec\sigma\right)/2$, which
is asymptotically reached by $\eta_t$ for every initial preparation of the atom:
\[
\vec{x}_\mathrm{eq} = -\gamma\Big(1-2c|\alpha_1|\sin(\vartheta_1-\varphi)\Big)\,A^{-1}\,
\begin{pmatrix}0\\0\\1\end{pmatrix}.
\]

\paragraph{Homodyne incoherent spectrum and fluorescence light squeezing.}
In a description of photo-detection based on quantum trajectories \eqref{posterior},
\eqref{currents}, the electromagnetic field has been traced out. Nevertheless, this
description is fully consistent with a model which includes also a quantum description of the
electromagnetic field and of its interaction with the atom and where $I_1$ and $I_2$ are the
outputs of measurements performed just on the emitted light \cite{BarPag96}. Therefore an
analysis of the homodyne photocurrents can reveal properties of the light detected in the
corresponding channels. In order to study the squeezing properties of the light in channel
$k$, the fundamental tool is the incoherent spectrum of $I_k$ \cite{BG08}
\begin{eqnarray*}
S_k(\mu)&=&\lim_{T\to+\infty}\frac{1}{T}\left\{\E\left[\abs{\int_0^T\rme^{\rmi\mu
s}\,I_k(s)\,\rmd
s}^2\right]\right. \\
&&{}-\left.\abs{\E\left[\int_0^T\rme^{\rmi \mu s}\,I_k(s)\,\rmd s\right]}^2\right\}.
\end{eqnarray*}
It is the limit of the normalized variance of the Fourier transform of the photocurrent $I_k$;
as usual, we call incoherent the part of the spectrum due to the fluctuations of the output.
The asymptotic behaviour of the atomic a priori state $\eta_t$ ensures that the limit exists.
It is a positive even function of its real argument $\mu$ which can be computed from equations
\eqref{posterior} and \eqref{currents} by Ito calculus and by the full theory of quantum
continual measurements, which can provide the first and second moments of $I_1$
\cite{Bar90QO,Bar05}. Thus, for every initial state of the atom, we get
\begin{equation}\label{spectrum}
S_k(\mu)=1+2\gamma|\alpha_k|^2\,\vec{s}_k\cdot\left(\frac{A}{A^2+\mu^2}\,\vec{t}_k\right),
\end{equation}
where $\vec{s}_k$ and $\vec{t}_k$ are the vectors in $\Real^3$ defined as
\begin{eqnarray*}
\vec{s}_k&=&\begin{pmatrix}\cos\vartheta_k,&\sin\vartheta_k,&0\end{pmatrix},\\
\vec{t}_k&=&\Tr\Big[\big(\rme^{\rmi\vartheta_k}\,\sigma_-\,\eta_\mathrm{eq} +
\rme^{-\rmi\vartheta_k}\,\eta_\mathrm{eq}\,\sigma_+
- \Tr[\sigma_{\vartheta_k}\,\eta_\mathrm{eq}]\,\eta_\mathrm{eq}\\
&&{}+ \delta_{k1}\,\frac{\rmi
c}{|\alpha_1|}\,[\eta_\mathrm{eq},\sigma_\varphi]\big)\,\vec\sigma\Big].
\end{eqnarray*}
Each spectrum $S_k$ depends on $\mu$, $k_\rmd$, $\overline{n}$, $\Omega$, $\Delta\omega$,
$\alpha_k$, $c$ and $\varphi$. Moreover, $S_2$ depends on $\alpha_1$, too. In the case $c=0$ (no feedback) each dependence on
$\varphi$ disappears and $S_2$ becomes independent of $\alpha_1$; then,
$S_1=S_2$ if $\alpha_1=\alpha_2$.

When $c=0$, for every $\mu$ and $\vartheta_k$, the value of $S_k(\mu;\vartheta_k)$ is the
variance of a quadrature of the light in channel $k$, the value of
$S_k(\mu;\vartheta_k+\pi/2)$ is the variance of the conjugate quadrature and Heisenberg-type
relations imply that $S_k(\mu;\vartheta_k)\,S_k(\mu;\vartheta_k+\pi/2)\geq1$ \cite{BG08}. The
light in channel $k$ is in a squeezed state if the variance of one quadrature is below the
standard quantum limit, that is if $S_k(\mu)<1$ for some $\mu$ and $\vartheta_k$.

When $c>0$, only the light in channel 2 is potentially available for homodyne detection with
arbitrary $\vartheta_2$, as well as it could be employed for different uses. On the contrary,
the light in channel 1 is a part of the feedback loop, it has to be detected, and a change
of $\vartheta_1$ implies a change of the atomic dynamics and, so, of the state of the emitted
light itself, not only a change of the quadrature under consideration. Of course, the spectrum
$S_1$ can be considered also in this case, but when $S_1(\mu)<1$ one can speak only of
`in-loop squeezing'. Its meaning and possible usefulness are discussed by Wiseman \cite{W98}.

For each channel we can give a measure of the `mean squeezing' of the light by introducing the
quantity
\begin{equation*}
\Pi_{k}(\vartheta_k)=\frac 1 {2\pi \gamma } \int_{-\infty}^{+\infty}\left[
S_{k}^{\mathrm{inel}}(\mu;\vartheta_k)-1\right] \rmd \mu= \abs{\alpha_k}^2 \,\vec t_k \cdot \vec{s}_k.
\end{equation*}
When $\Pi_{k}(\vartheta_k)<0$ for some $\vartheta_k$, the light in channel $k$ is surely
squeezed, but the spectrum can go below 1 even if $\Pi_{k}(\vartheta_k)$ is
positive. Moreover, we introduce the \emph{squeezing parameter} of the state of the light in
channel 2
\begin{equation}\label{linkalsq}
\Sigma_2=\inf_{\vartheta_2}\Pi_{2}(\vartheta_2)=
\abs{\alpha_2}^2\left[\mathrm{AS}_{\eta_{\mathrm{eq}}}+z_{\mathrm{eq}}+\abs{z_{\mathrm{eq}}}\right].
\end{equation}

\paragraph{Control of squeezing.}
We are interested in the squeezing properties of the fluorescence light and of the atom.
Regarding the fluorescence light, we can consider the squeezing of the light in the channels 1
and 2. Regarding the atom, we can consider the squeezing of the a posteriori state $\rho_t$
and of the a priori state $\eta_t$, and in particular of its limit $\eta_\mathrm{eq}$. Let us
stress that the definition of atomic squeezing does not depend on the fact that we are working
in the rotating frame. Let us start by investigating the effect of the control parameters.

Independently of the presence of the feedback loop,
every time a parameter $|\alpha_k|$
vanishes, the corresponding photocurrent $I_k$ reduces to a pure white noise (shot noise due
to the local oscillator) with spectrum $S_k=1$.

\emph{The case $c=0$.} In this case the dependence of each spectrum $S_k$ on the corresponding
$|\alpha_k|$ reduces to the explicit multiplication coefficient in \eqref{spectrum}.
Therefore, when the control parameters $\Omega$ and $\Delta\omega$ give squeezed light in
channel $k$, the lowering of $S_k$ under the shot noise level is anyhow directly proportional
to the fraction of emitted light gathered in that channel.

For $\Omega=0$ and $\overline n=0$ there is no
fluorescence light in the long run, so that each photocurrent $I_k$ asymptotically
reduces to a pure white noise and $S_k=1$.

For $\Omega=0$ and $\overline n>0$ there is no dependence on $\vartheta_k$ and $S_k>1$.
In this case there is only thermal light with carrier frequency $\omega_0$, while
the local oscillator is at frequency $\omega$. The result are two temperature
dependent Lorentzian peaks at $\mu=\pm \Delta \omega$.

\begin{figure}
\includegraphics[scale=.255]{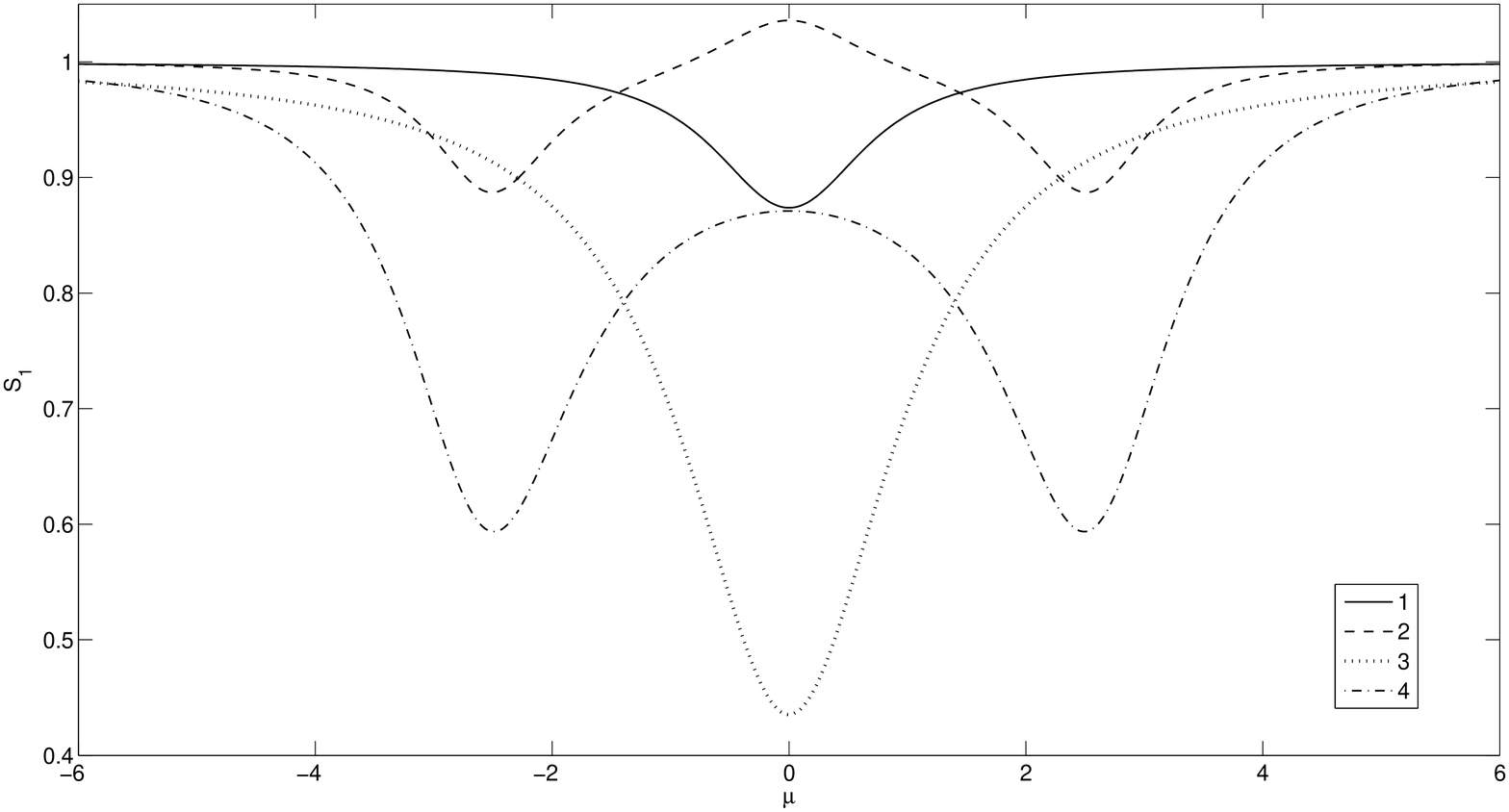}
\caption{\label{fig1} Channel 1}
\end{figure}

\begin{figure}
\includegraphics[scale=.255]{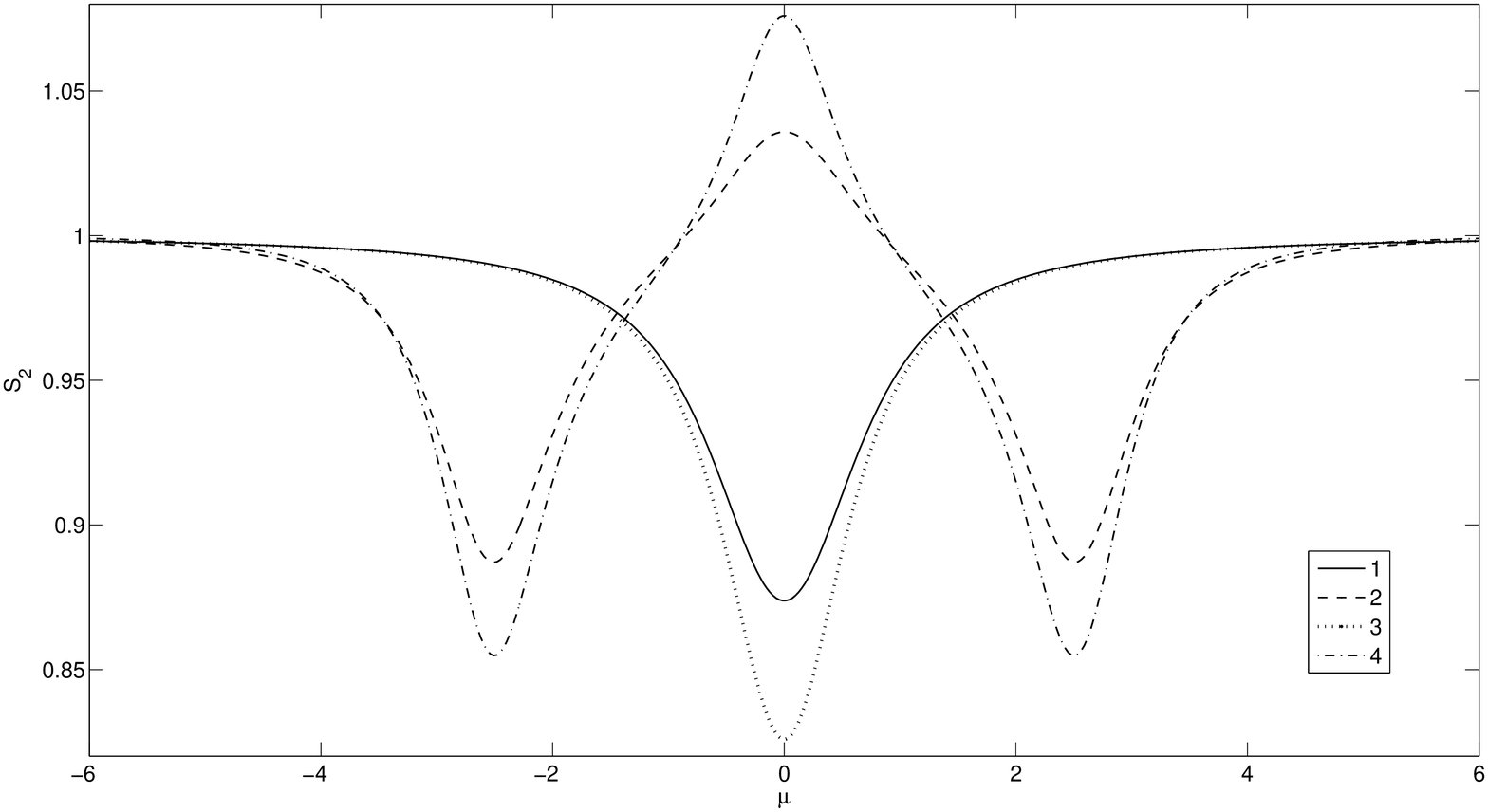}
\caption{\label{fig2} Channel 2}
\end{figure}

When $\Omega>0$, $S_k$ becomes $\vartheta_k$-dependent and it can go below the shot noise
level. This fact means that some negative correlation between the terms of the photocurrent
\eqref{currents} has been developed. Some examples are plotted for both channels. All the
figures are always in the case $\gamma=1$, $k_\rmd =\overline n=0$ and
$|\alpha_1|^2=|\alpha_2|^2=0.45$, with the control parameters used first to fix the position
of the minima of $S_k$ and, then, to have the lowest minima. Figures \ref{fig1} and \ref{fig2}
show $S_1$ and $S_2$ (which are equal in this case) for minima in $\mu=0$ (line 1, with
$\Delta\omega=0$, $\Omega=0.2976$, $\vartheta_k=-\pi/2$) and in $\mu=\pm 2.5$ (line 2, with
$\Delta\omega=1.8195$, $\Omega=1.7988$, $\vartheta_k=-0.1438$).

One could also compare the homodyne spectrum with and without $k_\rmd$ and $\overline n$, thus
verifying that the squeezing is very sensitive to any small perturbation.

Regarding the atom, $\mathrm{AS}_{\eta_{\mathrm{eq}}}$ depends only on $k_\rmd$,
$\overline{n}$, $\Omega$, $\Delta\omega$. In the case $k_\rmd=\overline n=0$, the
condition $\mathrm{AS}_{\eta_{\mathrm{eq}}}<0$ becomes $0<2\Omega^2 <   4\Delta\omega^2
+\gamma^2$ and the minimum value of $\mathrm{AS}_{\eta_{\mathrm{eq}}}$ is $-1/8$, reached for
$6\Omega^2 = 4\Delta\omega^2 +\gamma^2$.

\emph{The case $c\geq 0$.} The optimal squeezing in channel 1 is always found for $\Omega=0$
and the feedback loop is very helpful, giving good visible minima of $S_1$ also when
$|\alpha_1|$ is not close to 1. For example, FIG.~\ref{fig1} shows $S_1$ for minima in $\mu=0$
(line 3, with $\Delta\omega=0$, $\Omega=0$, $c=0.2936$, $\varphi-\vartheta_1=\pi/2$) and in
$\mu=\pm 2.5$ (line 4, with $\Delta\omega=2.5499$, $\Omega=0$, $c=0.3772$,
$\vartheta_1=-1.3354$, $\varphi=-0.0646$). The utility of the feedback scheme can be
appreciated by comparing line 1 with line 3 and line 2 with line 4.

If we are interested in the light emitted in channel 2 and if $|\alpha_1|$ and $|\alpha_2|$
are assigned by some constraints, then the squeezing in channel 2 can be enhanced by a
feedback scheme based on the photocurrent coming from channel 1, but the feedback performance
is not as good as it can be for the squeezing in channel 1 itself. FIG.~\ref{fig2} shows $S_2$
for minima in $\mu=0$ (line 3, with $\Delta\omega=0$, $\Omega=0.2698$, $\vartheta_1=\pi/2$,
$c=0.0896$, $\varphi=0$, $\vartheta_2=-\pi/2$) and in $\mu=\pm 2.5$ (line 4, with
$\Delta\omega=1.6920$, $\Omega=1.9276$, $\vartheta_1=2.8168$, $c=0.1326$, $\varphi=1.2460$,
$\vartheta_2=-0.0851$).

Anyway, if the only constraint is $|\alpha_1|^2 + |\alpha_2|^2 = 1-|\alpha_0|^2$ and we are
free in the choice of $|\alpha_1|$ and $|\alpha_2|$, then the best observable squeezing in
channel 2 is obtained in the case $|\alpha_1|=0$, $c=0$. That is, when the whole
non-lost light is gathered just in channel 2 and the white noise $I_1$ revealed in channel 1
is ignored.

The feedback loop can be really efficient also to enhance the atomic squeezing. For example,
in the ideal situation $|\alpha_1|=1$, $k_\rmd =\overline n=0$, with $\gamma=1$,
$\Delta\omega=3$, $\Omega=4$, $\vartheta_1=\pi/2$, $c=1.3372$, $\varphi=-\pi/40$, we get
$\mathrm{AS}_{\eta_{\mathrm{eq}}}=-0.2414$, which is very close to the bound $-1/4$. In this
case $\eta_\mathrm{eq}$ is almost pure so that also the a posteriori state $\rho_t$ is frozen
in a neighbourhood of $\eta_\mathrm{eq}$ and $\mathrm{AS}_{\rho_t}$ is maximized, too.

\paragraph{Fluorescence light vs atomic squeezing.}
There are not simple relations among the squeezing properties of fluorescence light in
channel 1, of fluorescence light in channel 2, of atomic a priori equilibrium state and of
atomic a posteriori state. Indeed, changing the parameters of our model, we can observe a
wide variety of behaviours.

The only clear link is the one mentioned above: if
$\mathrm{AS}_{\eta_{\mathrm{eq}}}\simeq-1/4$, then $\eta_\mathrm{eq}$ is almost pure and
$\rho_t$ is frozen in a neighbourhood of $\eta_\mathrm{eq}$, so that $\mathrm{AS}_{\rho_t}$ is
minimized, too, and the fluorescence light squeezing disappears as there is not incoherent
scattering of light. One can check that actually only the coherent scattering survives, giving
a $\delta$-contribution in $\mu=0$ to the complete spectrum. If the freezing of the atom is
only approximate, one can check that all the spectra tend to become flatter and the
fluorescence light squeezing tends to disappear.

There is also the link \eqref{linkalsq} between $\mathrm{AS}_{\eta_{\mathrm{eq}}}$ and
$\Sigma_2$. This gives a direct relation between atomic and fluorescence light squeezing in
absence of feedback. Indeed, in this case we have $z_\mathrm{eq}\leq0$, so that
$\Sigma_2=\abs{\alpha_2}^2\,\mathrm{AS}_{\eta_{\mathrm{eq}}}$, and we can consider the case
$|\alpha_1|=0$, so that $|\alpha_2|^2$ is the fraction of the whole detected light. This
relation is essentially the same found by Walls and Zoller considering a single mode for the
emitted light \cite{WallsZ81}. However the relation is not fundamental, as the feedback loop
can give $z_\mathrm{eq}>0$ and in this case we have always $\Sigma_2\geq0$ even if
$\mathrm{AS}_{\eta_{\mathrm{eq}}}<0$.

There is no relation between fluorescence light squeezing revealed in channel 1 and in channel
2, even if we fix the constraint $|\alpha_1|=|\alpha_2|$. For example, the lowest minima of
$S_1$ are found for $c>0$ and $\Omega=0$, but, every time $\Omega=0$, the light in channel 2
is not squeezed as it can be proved that $S_2\geq1$ for every $\mu$ and every $\vartheta_2$.

It is worth mentioning also the case $|\alpha_1|=1$, $\gamma=1$, $k_\rmd=\overline n=0$, with
$\Delta\omega=0$, $\Omega=0$, $\vartheta_1=\pi/2$, $c=1.2818$, $\varphi=0$. Then we have an
extremely visible squeezing in channel 1 ($S_1$ reaches 0.3183), there is no squeezing of the
atomic a priori equilibrium state ($\mathrm{AS}_{\eta_{\mathrm{eq}}}=0.0922$), while numerical
simulations show that the a posteriori state $\rho_t$ tends to become pure (as $|\alpha_1|=1$)
with $\mathrm{AS}_{\rho_t}$ stochastically moving between $-1/4$ and 0.

Finally let us remark that the idea of the papers \cite{WW00} is to choose the control
parameters in such a way that, in the rotating frame, the atom is frozen in a preassigned pure
state $h_0\in\Complex^2$, i.e.\ in such a way that, in the rotating frame, both the a priori
state $\eta_t$ and the a posteriori state $\rho_t$ asymptotically reach
$\eta_\mathrm{eq}=|h_0\rangle\langle h_0|$. This is possible in an exact way only in a very
ideal case, which in our notations corresponds to $|\alpha_1|=1$, $k_\rmd=\overline n=0$,
$\Delta\omega=0$, $\vartheta_1=\pm\pi/2$, $\varphi=0$, which implies in particular
$a_{12}=a_{21}=0$ and $x_\mathrm{eq}=0$. Then, $\rho_t$ is driven to a pure given state if
$\Omega$ and $c$ are such that $y_\mathrm{eq}^{\;2}+z_\mathrm{eq}^{\;2}=1$ and $2c \sin
\vartheta_1=1+z_\mathrm{eq}$. But this implies $\vec{t}_1=0$ and the two incoherent spectra
reduce to pure shot noise.


\vspace*{-6pt}


\begin{thebibliography}{99}
\bibitem{Dav76} E.~B.~Davies, \textit{Quantum Theory of Open Systems} (Academic, London,
    1976); A.~Barchielli, V.~P.~Belavkin, J. Phys. A: Math. Gen. \textbf{24}, 1495
    (1991).
\bibitem{Bar90QO} A.~Barchielli, Quantum Opt. \textbf{2}, 423
    (1990).
\bibitem{BarPag96} A.~Barchielli, A.~M.~Paganoni, Quantum Semiclass. Opt. \textbf{8},  133
    (1996).
\bibitem{Bar05} A.~Barchielli, in \textit{Open Quantum Systems III}, edited by S.~Attal,
    A.~Joye, C.-A.~Pillet, Lecture Notes in Mathematics \textbf{1882}, p.~207 (Springer, Berlin, 2006).
\bibitem{WisMil93} H.~M.~Wiseman, G.~J.~Milburn, Phys. Rev. A \textbf{47}, 1652 (1993).
\bibitem{WisM93} H.~M.~Wiseman, G.~J.~Milburn, Phys. Rev. Lett. \textbf{70}, 548 (1993);
    H.~M.~Wiseman, Phys. Rev. A \textbf{49}, 2133 (1994).
\bibitem{WW00} J.~Wang, H.~M.~Wiseman, Phys. Rev. A \textbf{64} (2001) 063810; J.~Wang,
    H.~M.~Wiseman, G.~J.~Milburn, Chemical Physics \textbf{268}, 221
    (2001); H.~M.~Wiseman, S.~Mancini, J.~Wang, Phys. Rev. A \textbf{66} (2002)
    013807.
\bibitem{WalM94} D.~F.~Walls, G.~J.~Milburn, \textit{Quantum Optics} (Springer, Berlin
    1994).
\bibitem{BGL08} A.~Barchielli, M.~Gregoratti, M.~Licciardo, arXiv:0801.4710.
\bibitem{WallsZ81} D.~F.~Walls, P.~Zoller, Phys. Rev. Lett.  \textbf{47}, 709 (1981).
\bibitem{W98} H.~M.~Wiseman, Phys. Rev. Lett. \textbf{81}, 3840 (1998).

\bibitem{BG08} A.~Barchielli, M.~Gregoratti, arXiv:0802.1877.

\end{thebibliography}
\end{document}